\documentclass[nofootinbib,aps,prd,preprint,groupedaddress,showpacs,showkeys]{revtex4-1}

\usepackage{amssymb}
\usepackage {amsmath}
\usepackage{natbib}
\usepackage{filecontents}
\usepackage{graphicx}
\usepackage{dcolumn}
\usepackage{bm}

\newcommand{\dif}{\mathrm{d}}

\begin{document}


\title{Modified entropies, their corresponding Newtonian forces, potentials, and temperatures}

\author{Aldo Mart\'inez-Merino}
 \email{a.merino@fisica.ugto.mx}
\author{Octavio Obreg\'on}%
 \email{octavio@fisica.ugto.mx}
\affiliation{%
 Departamento de F\'isica, Divisi\'on de Ciencias e Ingenier\'ia \\ Universidad de Guanajuato, Campus Le\'on \\ Loma del Bosque No. 103, Fracc. Lomas del Campestre, Le\'on, Guanajuato, M\'exico
 }%
\author{Michael P. Ryan Jr.}
 \email{ryanmex2002@yahoo.com; Permanent address: 30101 Clipper Lane, Millington MD 21651, USA}
 \affiliation{Instituto de Ciencias Nucleares, Universidad Nacional Aut\'onoma de M\'exico,\\A. Postal 70-543, M\'exico, D.F., M\'exico}



\date{\today}

\begin{abstract}
Assuming the hypothesis of the entropic nature of gravity, we calculate generalized Newtonian forces, their associated potentials and field equations, when other, in general non-extensive, entropies are considered instead of the usual Boltzmann entropy. Some comments about the nature of the correction terms presented are given, and a calculation of corrections to the Bekenstein-Hawking temperature is performed using the arguments which lead to what can be called a Michell-Laplace black hole.
\end{abstract}

\pacs{04.25.-g, 04.25.Nx, 04.50.-h, 04.50.Kd, 04.70.Dy}
\keywords{Newtonian potentials, entropy, entropic gravity}
\maketitle


\section{\label{sec:level1}Introduction}

\noindent One of the principal paradigms of our time is that of the fundamental forces in nature. We are taught that the electromagnetic, weak, strong and gravitational interactions control all natural phenomena. The Standard Model of particle Physics has made impressive achievements in unifying the first three into just one framework using the field concept. When this theory is quantized, the concept of particle arises, which is the basis of how we describe interactions. It is true that with quantum theory we can study systems like atoms or molecules, but the very core of the fundamental physics is field theory.\\
\indent For gravity the story has not been as successful. Out of all of these forces, gravity does not yet have a fully workable quantization. Even less is known about how to unify gravity with the other forces. A perturbative quantum version of Einstein's theory of General Relativity leads to divergences which are not renormalizable, so a complete quantum version of Einstein's theory is needed. With such a theory we could solve the black hole information paradox \cite{Hawking3}, and answer questions about the beginning of our Universe, that is, the behavior of spacetime singularities. However, what if gravity is not a fundamental force? In the physics community there is some speculation about whether this constant failure to fully quantize gravity is due to the fact that we are trying to quantize an effective theory.\\
\indent It is undeniable that General Relativity (GR) is an accurate theory which describes the dynamics of astronomical objects at scales from the solar system to clusters of galaxies. Also, the recent detection of gravitational waves by the LIGO \cite{LIGO} collaboration agrees well with the predictions of GR. Nevertheless, the question about whether GR is the ultimate theory is very relevant.\\
\indent The resemblance of the so-called laws of black holes with that of thermodynamics first noticed by J.D. Bekenstein \cite{Bekenstein}, along with many works of S. Hawking \cite{Hawking1, Hawking2, Hawking3} concerning the creation of particles by a gravitational field, or recently, the derivations of Einstein's equations from entropy by T. Jacobson et al. \cite{Jacobson1, Jacobson2, Jacobson3, Jacobson4}, suggest that GR may be an effective theory. It is worth mentioning the work of M. van Raamsdonk et al. \cite{VanRaamsdonk} where the Einstein equations are derived from the \textit{laws of entanglement}.\\
\indent All this having been said, the work of E. Verlinde \cite{Verlinde} on the derivation of  Newton's law of gravity using holographic arguments is in the same spirit as the above works. It is worth emphasizing that this derivation makes use of the Bekenstein-Hawking entropy-area relation, which characterizes black holes. Verlinde suggests that gravity is an entropic phenomenon, a force resulting from the response in changes of information entropy which is stored in some holographic sphere.\\
\indent Up to this point we have concentrated on General Relativity. However, it turns out that Newtonian gravity with a few modifications, along with some extremely simplified ideas from quantum field theory, allow us to obtain several important results from black hole theory. The authors have prepared a separate companion article \cite{MOR} where we discuss several uses of such an augmented Newtonian theory applied to black holes, calculating a ``Hawking'' temperature and the analog of the Bekenstein-Hawking entropy, giving a Newtonian version of Jacobson's \cite{Jacobson1} derivation of the Einstein field equations from entropy, and studying these ideas applied to a higher-derivative modified Newtonian gravity.\\
\indent Since it seems to be true that Newton's law is an entropic force, one should be able to modify it by considering more general entropies, and studying these modifications in order to gain insight into the corrections to GR that would give us these modifications in a Newtonian limit. Take, for example, the entropy derived from Loop Quantum Gravity (LQG) \cite{RovAsh}; the corrections generated by this entropy are logarithmic and given in powers of the inverse of the area. It is possible to calculate the potential producing such a force. We would like to be able to find the field equation related to this new generalized Newtonian gravitation, and give some interpretation of such terms. Such corrections may come from the weak field approximation of a modification of GR. That is, given some generalized entropy, we can study the correction terms arising from this modification \cite{ModestoRandono}.\\
\indent In this paper we attempt to find the field equations for modifications to Newton's law arising from these generalized entropies and calculate the corresponding Newtonian potentials. In order to see how far our results can be carried out, we also compute corrections to the Hawking temperature, by making the same assumptions that lead to what we call a Michell-Laplace black hole. That is, by means of classical arguments, we can compute the radius of an object whose escape velocity is that of light. With this result, coupled with some extremely simplified concepts from quantum field theory, we can calculate the``Hawking'' temperature of such an object by comparing its thermal energy with its total energy.\\
\indent Finally, we will use the calculation of an entropy given by a simple calculation of a black hole entropy using the Clausius relation with the ``Hawking'' temperature. We can then compare this result with the usual entropy. This calculation leads to an interesting conundrum with respect to the entropy-area relation of these theories.

\section{Quantum modifications to Newton's Law}

\noindent We begin by reviewing the entropies we consider in this paper. As Bekenstein first noted, entropy is a concept linked to the geometry of some characteristic of the system. In the case of the black hole, entropy is related to the area of its event horizon, given by the expression $\mathcal{S}_{BH} = \frac{A}{4 l_p^2}$, a formula due to Bekenstein and Hawking. It is known how to obtain corrections to this entropy that come from quantum gravity effects. It can be done by means of a path integral approach\footnote{In \cite{LOSR}. The authors perform the calculation considering noncommutativity in such black hole models.} \cite{OST}, or the trace anomaly \cite{Fursaev}, or counting spin states in Loop Quantum Gravity (LQG) \cite{RovAsh}. The expression is given by
\begin{equation}
\mathcal{S} = \frac{A}{4 l_p^2} - \beta \mathrm{ln} \frac{A}{4 l_p^2} + \gamma \frac{4 l_p^2}{A}; \label{LQGe}
\end{equation}

\noindent where $\beta$ and $\gamma$ are very small parameters. The logarithmic correction arises as a consequence of several approaches.\\
\indent The relation between entropy and the geometry of the system is exploited by Jacobson in his derivation of Einstein's equations \cite{Jacobson1}, where he made use of Clausius relation $\delta E = T \delta S$, and where the matter present is considered as a part of the energy. Verlinde \cite{Verlinde} uses this concept to derive the Newtonian law of gravitation. He interprets the entropy as information concerning the position of material bodies around a point mass $M$ at a distance $R$; all points at this distance define a sphere $\mathbb{S}$ embedded in a three-dimensional space. Thus when we change the bits of information stored on the surface, a force appear as a reaction to that change. Choosing some constants, and assuming a linear relation between the quantity of bits stored and the area of such a sphere, the resulting entropic force is (note that we are assuming spherical coordinates)
\begin{equation}
\mathbf{F} = - \frac{GMm}{R^2} 4 l_p^2 \frac{\partial \mathcal{S}}{\partial A} \Big|_{A = 4 \pi R^2} \hat{\mathbf{R}}, \label{EntropicForce}
\end{equation}

\noindent where $\hat{\mathbf{R}}$ is a unit radial vector. We arrive at the Newtonian force when we consider the Bekenstein-Hawking entropy, $\mathcal{S}_{BH}$. Note that the expression (\ref{EntropicForce}) is general enough to allow us to consider different kinds of entropies. In fact, in \cite{Sheykhi} the author uses the entropy (\ref{LQGe}) to derive a Newtonian cosmology through some modification of the Friedmann equations\footnote{We also cite \cite{Frampton} for a study of entropic correction terms in cosmology.}. The resulting force associated with the entropy (\ref{LQGe}), is
\begin{equation}
\mathbf{F} = - \frac{GMm}{R^2} \left[ 1 - \beta \frac{l_P^2}{\pi R^2} - \gamma \left( \frac{l_P^2}{\pi R^2} \right)^{2} \right] \hat{\mathbf{R}}. \label{LQGForce}
\end{equation}

\noindent In particular for this entropy, the correction terms are quantum in nature, provided that the entropy is already quantum. As pointed out in \cite{ModestoRandono}, these correction terms are compatible with corrections coming from the area operator and the volume element in LQG.\\
\indent At this stage we want to stress some more points about this force. It is well known that Newtonian gravitation is recovered from Einstein's General Relativity in the weak field approximation. What is recovered is the field equation for some matter density, $\rho$,
\begin{equation}
\nabla^2 \Phi = 4 \pi \rho, \label{FieldEq}
\end{equation}

\noindent where $\Phi$ is the potential function that gives rise to the force, and $\nabla^2$ is the three-dimensional, flat-space Laplacian. The potential function for the force (\ref{LQGForce}) can be computed straightforwardly \cite{ModestoRandono},
\begin{equation}
\Phi = -\frac{GM}{R} \left( 1 - \frac{\beta l_P^2}{3 \pi R^2} - \frac{\gamma l_P^4}{5 \pi^2 R^4} \right), \label{qPotential}
\end{equation}

\noindent and

\begin{equation}
\nabla^2 \Phi = -\frac{\nabla \cdot \mathbf{F}}{m} = 2 \frac{GM}{R} \frac{l_P^2}{\pi R^4} \left[ \beta + 2\gamma \frac{l_P^2}{\pi R^2} \right]. \label{qFieldEq}
\end{equation}

\indent Now, given $\Phi$ we would like to go further and find a modified Newtonian field equation that would have the $\Phi$ of Eq. (\ref{qPotential}) as its solution. This, of course, is difficult. It is like asking, given an ordinary function $y = x^3$, what differential equation would have such a function as its solution. Without some restriction on the differential equation there are an infinite number of answers. In the case of Eq. (\ref{qFieldEq}) one possibility is to simply assume that (\ref{FieldEq}) is valid with $\rho$ some effective matter density, $\rho_{\mbox{eff}}$, with
\begin{equation}
\rho_{\mbox{eff}} = \frac{GM}{2 \pi R} \frac{l_P^2}{\pi R^4} \left[ \beta + 2 \gamma \frac{l_P^2}{\pi R^2} \right],
\end{equation}

\noindent where $\rho_{\mbox{eff}}$ would be due to some sort of ``quantum foam'' generated by the existence of a point mass $M$  by means of an unknown quantum process. Another, somewhat more conventional philosophy, would be to assume that $\Phi$ is generated from a point mass $M$ using some higher-order Newtonian limit of an unknown quantum gravity theory where the matter density $\rho$ is zero outside the point mass.\\
\indent As an exercise, the simple form of $\Phi$ (a sum of functions of the form $R^{-n}$) allows us to mock up one possibility of an ad hoc field equation with only a point source on the right-hand-side. There is certainly no guarantee that this procedure would give a desirable field equation. However, one such equation can be cobbled together by writing
\begin{equation}
\Phi = H(R) \Phi_0(R),
\end{equation}

\noindent $\Phi_0$ the potential above and $H(R)$ the Heaviside step function,
\begin{eqnarray}
H(R) = \left\{
\begin{array}{ll}
1 & R>0, \\
1/2 & R=0, \\
0 & R<0.
\end{array} \right.
\end{eqnarray}

\noindent With some numerical juggling the following combination gives a delta function point source for $\Phi = \Phi (R)$ (notice that the radial direction $\hat{\mathbf{R}}$ appears, so this equation is only valid for spherically symmetric potentials),
\begin{equation}
\nabla^2 \Phi + \kappa R^2 \nabla^2 (\nabla^2 \Phi) + \alpha \left( \frac{\nabla \Phi \cdot \hat{\mathbf{R}}}{R} + \frac{\Phi}{R^2} \right)
\end{equation}

\noindent with $\kappa = -1/75$ and $\alpha = 11/5$. We now have [with $\Phi_0$ as given in (\ref{qPotential})],
\begin{equation}
\nabla^2 \Phi_0 + \kappa R^2 \nabla^2 (\nabla^2 \Phi_0) + \alpha \left( \frac{\nabla \Phi_0 \cdot \hat{\mathbf{R}}}{R} + \frac{\Phi_0}{R^2} \right) = 0.
\end{equation}

\noindent The final equation becomes (using $\dif H(R)/ \dif R = \delta(R)$ and the well-known relations $\dif \delta (R) / \dif R = - \delta(R) / R$ and $\delta (\mathbf{R}) = \delta (R) / 4 \pi R^2$)
\begin{eqnarray}
\nabla^2 \Phi &+& \kappa R^2 \nabla^2 (\nabla^2 \Phi) + \alpha \left( \frac{\nabla \Phi \cdot \hat{\mathbf{R}}}{R} + \frac{\Phi}{R^2} \right) \nonumber \\
&=& 4 \pi GM \delta (\mathbf{R}) \left[-\frac{92}{75} - \frac{68}{75} \frac{\beta l_P^2}{3 \pi R^2} + \frac{783}{75} \frac{\gamma l_P^4}{5 \pi^2 R^4} \right]. \label{laplacian1}
\end{eqnarray}

\noindent The right-hand-side of this equation is rather clumsy and quite singular, which indicates that it is probably not a desirable field equation. However, Gauss's law implies that an integral of $\nabla \Phi$ over a sphere of radius $a$ has terms in negative powers of $a$, making a simple point mass source $\rho = GM \delta (\mathbf{R})$ impossible. As we mention in the companion article, entropies of this sort may possibly be used to find field equations using the results of Jacobson et al. \cite{Jacobson1, Jacobson2, Jacobson3, Jacobson4}, that are consistent with modified entropies.\\
\indent We would like to stress that under the assumption of this entropic nature of gravity, field equations derived from quantum corrections to the entropy would correspond to an associated generalized gravity whose weak field approximation must be this field equation. This may give some insight into possible terms that correct the Einstein equations in the quantum realm. For example, consider the Born-Oppenheimer approximation as applied by C. Kiefer \cite{Kiefer}, where the terms may be manifestations of heavy degrees of freedom, as they have been decoupled from the lighter ones which come from matter fields. At the present we do not have any idea of what such terms might be in the modified theories of gravity, but we are pursuing the matter at what we assume to be the linearized level.\\
\indent As this example teaches us, Eq. (\ref{EntropicForce}) should be general enough to allow us to consider other entropies, and lead us to find the corresponding potential functions and entropic forces. We note the appearance of the Planck length as a characteristic parameter. We are now going to consider other entropies very different from the one we have just discussed.

\section{Entropies derived from superstatistics}

\noindent One of the main characteristics of Boltzmann entropy is that it is extensive, the sum of the entropies for any two parts of a given system is the entropy of the system as a whole. Generalizing this entropy comes with the drawback that this property is, in general, lost. However, non-extensive entropies seem to be rewarding examples of more general phenomena. In Ref. \cite{BeckCohen} the authors treat a broad range of entropies using different Boltzmann factors, which have been called \textit{superstatistics}, obtained from different temperature distributions. In particular, we consider the cases of the Tsallis entropy \cite{Tsallis1}, and the one reported in \cite{Obregon1}; (see also \cite{Obregon2}).\\
\indent There are many examples where the Tsallis entropy appears, ranging from statistical systems to particle physics \cite{Tsallis2}. It relies heavily on a free parameter that takes different values for each system. The entropy reduces to Boltzmann's when this parameter is one. The other example is an entropy that depends solely on the probability, and differs from the Boltzmann entropy for large probabilities when the states of the systems are not large enough. A recent review of this entropy and some of its features is in \cite{Obregon3}. In \cite{ObregonCabo}, the authors discuss the discrepancy for a 2d CFT. They calculate the corrections to a 2dCFT and show that they correspond to an $AdS_3$ length-dependent entanglement entropy \cite{RT}.\\
\indent For now, let us explicitly give the functional form of these entropies\footnote{Note that in the expressions we are dealing with dimensionless quantities, we are already dividing the entropy by the Boltzmann constant, $k_B$.}:
\begin{itemize}
\item For the Tsallis entropy
\begin{equation}
\mathcal{S}_q = \frac{1}{q - 1} \left( 1 - \sum_{l=1}^{\Omega} p_l^{q} \right). \label{tsal}
\end{equation}

\item For the entropy of Obreg\'on\footnote{There is yet another entropy of the form $\mathcal{S} = \sum_{l=1}^{\Omega} \left( p_l^{-p_l} - 1\right)$ which might be of interest.}
\begin{equation}
\mathcal{S} = \sum_{l=1}^{\Omega} \left( 1 - p_l^{p_l}\right).
\end{equation}
\end{itemize}

\noindent From Eq.(\ref {tsal}) is easy to see that when $q \rightarrow 1$ we recover the Boltzmann entropy for Tsallis case, and for the Obreg\'on case we have the same Boltzmann entropy as a first term in the expansion.\\
\indent In order to calculate the potential function of the forces these entropies produce, we need to express these same entropies in a convenient form. Consider the special case of equiprobability, assuming a number $\Omega$ of states; that is, $p_l = \frac{1}{\Omega}$. Furthermore, we associate the Boltzmann entropy with the Bekenstein-Hawking entropy for a black hole, linear in the black hole area. As a result, the entropies we are going to use take the following forms
\begin{eqnarray}
\mathcal{S}_q &=& \frac{1}{q-1} \left(1 - e^{-(q-1) \mathcal{S}_{BH}} \right), \\
\mathcal{S} &=& e^{\mathcal{S}_{BH}} \left( 1 - e^{- \mathcal{S}_{BH} e^{-\mathcal{S}_{BH}}}\right). \label{OEntropy}
\end{eqnarray}

\noindent These entropies will then depend on the area through the Bekenstein-Hawking formula. Both the LQG entropy and the entropy of (\ref{OEntropy}) can be written in the form
\begin{equation}
\mathcal{S} = \mathcal{S}_{BH} + \mathfrak{s}(A),
\end{equation}

\noindent with the term $\mathfrak{s}$ including the corrections to the BH entropy relation. (We will consider the Tsallis entropy later, since it has a different form.)\\
\indent For these two entropies the associated forces are Eq. (\ref{LQGForce}), and for (\ref{OEntropy}) (expanded to first order in the doubled potential) are
\begin{equation}
F = - \frac{GMm}{R^2} \left[ 1 - \left( 1 - \frac{1}{2} \frac{\pi R^2}{l_P^2} \right) \frac{\pi R^2}{l_P^2} e^{-\frac{\pi R^2}{l_P^2}} \right] \hat{\mathbf{R}}.
\end{equation}

\indent The potential function for this entropic force is easily calculated. For the LQG force we have (\ref{qPotential}), and for the entropy (\ref{OEntropy}),
\begin{eqnarray}
\Phi = -\frac{GM}{R} + \frac{3}{8} \frac{GM \pi}{l_P} \mathrm{erfc} \left(\frac{\sqrt{\pi} R}{l_P} \right) - \frac{GM \pi}{4 l^2_P} R e^{-\frac{\pi R^2}{l_P^2}}. \label{OPotential}
\end{eqnarray}

\noindent In what follows we will be interested in values of $R$ large compared to $l_P$, so the argument of the error function will be very large, so we can use its asymptotic expansion, the first term of which is $\mathrm{erfc}(z) \cong (1/\sqrt{\pi} z) e^{-z^2}$, and
\begin{equation}
\Phi \cong -\frac{GM}{R} \left( 1 - \frac{3}{8} e^{-\frac{\pi R^2}{l_P^2}} + \frac{\pi}{4 l_P^2} R^2 e^{-\frac{\pi R^2}{l_P^2}}\right). \label{OPotentialA}
\end{equation}

\noindent Notice that both of these potentials (\ref{qPotential}) and (\ref{OEntropy}) are of the form
\begin{equation}
\Phi (R) = -\frac{GM}{R} [1 + \Phi_1 (R)], \label{GenPot}
\end{equation}

\noindent $\Phi_1$ very small for large $R$.\\
\indent In our companion article we have studied this class of potentials in several situations that will be relevant in the next section. One other equation we will need is the force associated with potentials of this form, that is, $\mathbf{F} = -m \nabla \Phi$,
\begin{equation}
\mathbf{F} = -\frac{GMm}{R^2} \left[ 1 + \Phi_1 (R) - R \frac{\dif \Phi_1(R)}{\dif R} \right] \hat{\mathbf{R}}. \label{force}
\end{equation}

\indent As previously, we can either consider the field equation for $\Phi$ to be
\begin{equation}
\nabla^2 \Phi = 4 \pi \rho_{\mathrm{eff}},
\end{equation}

\noindent where in the case of Eq. (\ref{OPotential}),
\begin{equation}
\rho_{\mathrm{eff}} = \frac{GM}{4 \pi R} \left[ - \left( \frac{\pi R^2}{l_P^2} \right)^2 - \frac{4 \pi R^2}{l_P^2} + 2 \right] e^{-\frac{\pi R^2}{l_P^2}}.
\end{equation}

\noindent Here any attempt to find a modified field equation for the potential (\ref{OPotentialA}) whose right-hand-side is zero by simply guessing is essentially impossible, so we will not attempt it.\\
\indent Returning to the case of the Tsallis entropy, we will see that it is not useful to expand the exponential because we expect the exponent in the resulting force to be very large and negative. We now have, replacing $\mathcal{S}_{BH}$ by $A/4l_P^2$,
\begin{equation}
\mathcal{S}_q = \frac{1}{q - 1} (1 - e^{-(q-1) A/4 l_P^2}).
\end{equation}

\noindent Now, $\dif \mathcal{S}_q / \dif A$ is simple and the resulting force from Eq. (\ref{EntropicForce}) is
\begin{equation}
\mathbf{F} = -\frac{GMm}{R^2} e^{-(q-1) \frac{\pi R^2}{l_P^2}} \hat{\mathbf{R}}.
\end{equation}

\noindent The associated potential is
\begin{equation}
\Phi = -\frac{GM}{R} \left( -\pi \sqrt{q-1} \frac{R}{l_P} \mathrm{erfc}\left[\sqrt{\pi} \sqrt{q-1}\frac{R}{l_P}\right] + e^{-(q-1) \frac{\pi R^2}{l_P^2}} \right).
\end{equation}

\noindent Unfortunately, if we use the asymptotic form of $\mathrm{erfc}$ we used in (\ref{OPotentialA}) the two terms in $\Phi$ cancel, so we must take the second term in its asymptotic series, $-e^{-z^2}/2\sqrt{\pi} z^3$, and
\begin{equation}
\Phi \cong - \frac{GMl_P^2}{2\pi (q-1) R^3} e^{-(q-1)\frac{\pi R^2}{l_P^2}}. \label{TsallisApprox}
\end{equation}

\noindent If we try to expand $\Phi$ to the first two terms in the exponential, we find the resulting $\Phi$ becomes positive for very large $R$ and the force becomes repulsive. It is best to leave this $\Phi$ as it is.\\
\indent We can calculate $\nabla^2 \Phi$ to find a $\rho_{\mathrm{eff}}$ directly as $-\nabla \cdot (\mathbf{F}/m)$, but unfortunately $\rho_{\mathrm{eff}}$ turns out to be negative. However, the very simple form of $\Phi$ does allow us to find a reasonable ad hoc field equation with zero on the right-hand-side. We have
\begin{equation}
\nabla^2 \Phi = -\frac{2\pi (q-1)}{l_P^2} \frac{GM}{R} e^{-(q-1) \frac{\pi R^2}{l_P^2}},
\end{equation}

\noindent and
\begin{equation}
\nabla^2 \Phi + \frac{\pi (q-1) R}{l_P^2} \nabla \Phi \cdot \hat{\mathbf{R}} = 0. \label{laplacian2}
\end{equation}

\noindent This equation still singularizes the radial direction, so it is also only applicable to spherically symmetric potentials.

\section{Corrections to the temperature of a black hole}

\noindent As an application of previous results, we calculate corrections to the temperature of a black hole by a non-orthodox method. Since we are working in a Newtonian framework, we compute such temperature using the assumptions that lead us to what in our companion article \cite{MOR} we call a Michell-Laplace black hole, which is a massive object whose escape velocity is equal to the speed of light. In our derivation, we are going to make use of the potential functions we have given in the previous section.\\
\indent Although this is a non-relativistic argument, we can obtain a rough estimate of the corrections to the temperature for the black hole. Consider a spherical object of mass $M$. Using conservation of energy, we can calculate the radius of this object must have in order for its escape velocity to be the speed of light. Considering the sum of the kinetic and potential energy of a projectile of mass $m$, we have
\begin{equation}
\frac{E}{m} = \frac{1}{2} \left( \frac{\dif R}{\dif t} \right)^2 - \frac{GM}{R}. \label{EnergyPerMass}
\end{equation}

\noindent If at an initial radius $R_0$, $\dif R/\dif t = c$ and the particle arrives at infinity with zero velocity,
\begin{equation}
R_0 = R_{ML} = 2GM / c^2.
\end{equation}

\noindent In our companion article we call $R_{ML}$ the \textit{Michell-Laplace radius} since it was first calculated independently by Michell \cite{Michell} and Laplace \cite{Laplace} in the 18th century. For the Newtonian potential, consider the production of a pair of virtual particles of mass $m$ and $-m$, at a distance $R = R_{ML} + l_c /2$, $l_c$ the Compton wavelength of the mass $m$. We assume that for some reason the negative mass particle always falls into the black hole (otherwise there would be no black hole evaporation) and the other moves radially away with velocity c. Using Eq. (\ref{EnergyPerMass}), the energy of the particle is
\begin{equation}
\frac{E}{m} \Big|_{R = R_{ML} + l_c / 2} = \frac{c^2}{2} - \frac{GM}{R_{ML} + l_c / 2} \cong \frac{c^2}{2} - \frac{GM}{R_{ML}} \left( 1 - \frac{l_c}{2R_{ML}} \right).
\end{equation}

\noindent In the last expression we assumed $l_c / R_{ML} <<1$ in order to perform the approximation. By conservation of energy, comparing this result with the energy of the particle at infinity, $E_{R = \infty} \equiv E_{\infty}$, we get
\begin{eqnarray}
E_{\infty} &=& \frac{1}{2} m c^2 - \frac{GMm}{R_{ML}} + \frac{\hbar c^3}{8 GM} \\
&=& \frac{1}{2} m c^2 - \frac{GMm}{2GM / c^2} + \frac{\hbar c^3}{8GM} \\
&=& \frac{\hbar c^3}{8 GM}.
\end{eqnarray}

\noindent We take this energy to correspond to a thermal energy of $T$ ($k_B = 1$), which gives a temperature of
\begin{equation}
T = \frac{\hbar c^3}{8 G M},
\end{equation}

\noindent which, up to a constant, is the Bekenstein-Hawking temperature. In our companion article we call this the \textit{Michell-Laplace-Hawking temperature}, $T_{MLH}$.\\
\indent In addition, we can compute the entropy of this system starting with the infall of a mass $\dif M$ into the black hole with velocity $c$. The increment in energy is $\dif M c^2 /2$, and using the Clausius relation $\dif E = T_{MLH} \dif \mathcal{S}$, we have
\begin{equation}
\frac{1}{2} \dif M c^2 = \frac{\hbar c^3}{8 GM} \dif \mathcal{S}, \label{Clausius}
\end{equation}

\noindent which upon integration yields,
\begin{equation}
\mathcal{S} = \frac{2 GM^2}{\hbar c}.
\end{equation}

\noindent This is the Bekenstein-Hawking entropy, $\frac{4 \pi GM^2}{\hbar c}$, up to a constant factor.\\
\indent As we said before, by this line of reasoning we are able to get a rough estimate of the temperature of a black hole, leading us to ask what the changes to this temperature would result if we used the above potentials associated with the entropies we are considering.\\
\indent In \cite{MOR}, we considered $\Phi$'s of the form (\ref{GenPot}), where $\Phi_1 (R_{ML})$ is very small. Since
\begin{equation}
\frac{1}{2} c^2 - \frac{GM}{R_{ML}} [1 + \Phi_1 (R_{ML})] = 0, \label{MLCondition}
\end{equation}

\noindent we have
\begin{equation}
\frac{2 GM}{c^2} \cong R_{ML} [1 - \Phi_1(R_{ML})],
\end{equation}

\noindent and assuming that $R_{ML} = 2GM/c^2 + \lambda$, $\lambda$ small, it is easy to show that
\begin{equation}
R_{ML} \cong \frac{2GM}{c^2} [1 + \Phi_1(2GM/c^2)].
\end{equation}

\noindent Now, for a particle of mass $m$ having $R_{ML} + l_c /2$ at velocity $c$,
\begin{equation}
E_{\infty} = \frac{1}{2} mc^2 - \frac{GMm}{R_{ML} + l_c/2} [1 + \Phi_1(R_{ML} + l_c/2)] = T_{MLH}.
\end{equation}

\noindent Expanding to first order in $l_c$, and using (\ref{MLCondition}),
\begin{eqnarray}
E_{\infty} &=& \frac{1}{2} Mc^2 \left( \frac{l_P}{R_{ML}} \right)^2 \left[ 1 + \Phi_1(R_{ML}) - R_{ML} \frac{\dif \Phi_1}{\dif R_{ML}}(R_{ML}) \right] \nonumber \\
&=& T_{MLH}.
\end{eqnarray}

\noindent Using $M = (c^2/2G) R_{ML} [1 - \Phi_1(R_{ML})]$, we finally have
\begin{equation}
T_{MLH} = \frac{1}{2} M_P c^2 \left( \frac{l_P}{R_{ML}} \right)^2 \left[ 1 - R_{ML} \frac{\dif \Phi_1}{\dif R_{ML}}(R_{ML}) \right]
\end{equation}

\noindent ($M_P$ the Planck mass). It will be useful below to have $T_{MLH}$ as a function of $M$, so inserting $R_{ML}$ as a function of $M$, we find
\begin{equation}
T_{MLH} = \frac{M_P c^2}{8M} \left[ 1 - \Phi_1(2GM/c^2) - M \frac{\dif \Phi_1(2GM/c^2)}{\dif M} \right].
\end{equation}

\indent For the LQG entropy we have
\begin{eqnarray}
T_{MLH} &=& \frac{1}{2} M_P c^2 \left(\frac{l_P}{R_{ML}}\right) \left[1 - \frac{2}{3} \frac{\beta l_P^2}{\pi R_{ML}^2} - \frac{4}{5} \frac{\gamma l_P^4}{\pi R_{ML}^4} \right] \nonumber \\
&=& \frac{M_P c^2}{8M} \left[1 - \frac{\beta}{12\pi} \left(\frac{M_P}{M}\right)^2 - \frac{3\gamma}{80 \pi} \left( \frac{M_P}{M} \right)^4 \right],
\end{eqnarray}

\noindent and for the entropy (\ref{OEntropy}) where $\Phi_1 = -\frac{3}{8} e^{-\frac{\pi R^2}{l_P^2}} + (\pi R^2 / 4l_P^2) e^{-\frac{\pi R^2}{l_P^2}}$,
\begin{eqnarray}
T_{MLH} &=& \frac{1}{2} \left(\frac{l_P}{R_{ML}}\right) \left[1 - \frac{3\pi}{4} \left(\frac{R_{ML}}{l_P}\right)^2 e^{-\frac{\pi R_{ML}^2}{l_P^2}} + \frac{\pi^2}{2} \left(\frac{R_{ML}}{l_P}\right)^4 e^{-\frac{\pi R_{ML}^2}{l_P^2}} \right] \nonumber \\
&=& \frac{M_P^2 c^2}{8M} \left[1 + \frac{3}{8} e^{-\pi (M/M_P)^2} - \pi \left(\frac{M}{M_P}\right)^2 e^{-\pi (M/M_P)^2} + \pi^2 \left( \frac{M}{M_P} \right)^4 e^{-\pi (M/M_P)^2} \right]. \nonumber \\
\end{eqnarray}

\indent For the Tsallis entropy, using (\ref{TsallisApprox}), we have
\begin{equation}
\frac{2GM}{c^2} = \frac{2\pi (q-1) R_{ML}^3}{l_P^2} e^{-\pi (q-1)\frac{R_{ML}^2}{l_P^2}}.
\end{equation}

\noindent Unfortunately, there seems to be no way to approximate this transcendental equation to give an approximate analytic solution for $R_{ML}$ as a function of $M$. However, if we write $R_{ML} = \alpha l_P$, we find
\begin{equation}
\frac{M}{M_P} = \pi (q-1) \alpha^3 e^{-\pi (q-1) \alpha^2},
\end{equation}

\noindent so even for large masses, $R_{ML}$ is essentially Planckian.\\
\indent If, notwithstanding, we formally expand $m \Phi_1 (R_{ML} + l_c/2)$ to first order in $l_c$, we have
\begin{eqnarray}
E_{\infty} &=& \frac{1}{2} M_P c^2 \left[-\frac{3}{2} \frac{l_P}{R_{ML}} + \frac{\pi (q-1)}{l_P} R_{ML} e^{-\pi (q-1) \frac{R_{ML}^2}{l_P^2}} \right] \nonumber \\
&=& T_{MLH}.
\end{eqnarray}

\noindent Without an analytic solution for $R_{ML}$ as a function of $M$, we cannot find $E_{\infty}(M)$ easily.

\section{Closing the circle}

\noindent An interesting check of the concepts that inform this article and its companion \cite{MOR} can be called ``closing the circle.'' In this article we began with several general definitions of entropy written as a function of $\mathcal{S}_B$, the Boltzmann entropy. Identifying $\mathcal{S}_{BH}$ with a black hole area, $\mathcal{S}_B = A/4l_P^2$, and then based on the work of Verlinde we constructed modified Newtonian forces by replacing $A$ with $4\pi R^2$.\\
\indent Now, once we have a modified Newtonian force and its associated potential $\Phi (R)$, we can use the results of Sec. 4 to calculate a Michell-Laplace black hole radius, $R_{ML}$, and a Michell-Laplace-Hawking temperature, $T_{MLH}$, of the black hole. We can now close the circle by using the idea of Eq. (\ref{Clausius}) to calculate an entropy $\mathcal{S}$ as a function of $R_{ML}$ or $M$, the black hole mass. This entropy can be written as a function of the black hole area, $A = 4\pi R_{ML}^2$, and it should be the same as the initial entropy for each one of the cases considered.\\
\indent We will use the entropies where $\Phi(R) = (-GM/R)[1 + \Phi_1(R)]$ to test this hypothesis. From Eq. (\ref{Clausius}) we have
\begin{equation}
\dif \mathcal{S} = \frac{\dif M c^2}{2T_{MLH}}.
\end{equation}

\noindent For reasons that will become clear below, we will express $\mathcal{S}$ both as a function of $R_{ML}$ and $M$. First substituting $M(R_{ML})$ to find $\dif M = [\dif M (R_{ML})/\dif R_{ML}] \dif R_{ML}$, we have
\begin{equation}
\frac{\dif \mathcal{S}}{\dif R_{ML}} = \frac{R_{ML}}{l_P^2} [1 - \Phi_1(R_{ML})], \label{difEntropy}
\end{equation}

\noindent so
\begin{equation}
\mathcal{S} = \frac{1}{2} \left[\frac{R_{ML}^2}{l_P^2} - 2 \int \frac{R_{ML}}{l_P^2} \Phi_1 \dif R_{ML} \right]. \label{Entropy1}
\end{equation}

\noindent As a function of $M$ we have
\begin{eqnarray}
\frac{\dif M c^2}{T_{MLH}} &=& \frac{4M}{M_P c} \left[1 + \Phi_1(2GM/c^2) + M \frac{\dif \Phi_1}{\dif M}(2GM/c^2)\right] \dif M \nonumber \\
&=& \dif \mathcal{S}.
\end{eqnarray}

\noindent Integration by parts gives us
\begin{equation}
\mathcal{S} = 2 \left(\frac{M}{M_P}\right) \left[1 + 2\Phi_1(2GM/c^2) -\frac{2}{M_P^2} \int M \Phi_1 (2GM/c^2) \dif M \right]. \label{Entropy2}
\end{equation}

\noindent We now want to ``close the circle'' by writing $\mathcal{S}$ as a function of area with the original $\mathcal{S} (\mathcal{S}_B)$ with $\mathcal{S}_B$ replaced by $A/4 l_P^2$. Obviously the black hole area is $4 \pi R_{ML}^2$. To avoid the integrals in (\ref{Entropy1}) and (\ref{Entropy2}) we would like to compare $\dif \mathcal{S} / \dif A$ with $A$ replaced by $4 \pi R^2$ with what we found in Eq. (\ref{force}). For $\mathcal{S}$ as a function of $R_{ML}$, we have $R_{ML} = \sqrt{A}/2\sqrt{\pi}$, and from (\ref{difEntropy}), we have
\begin{equation}
\mathbf{F} = -\frac{1}{8 \pi} \frac{GMm}{R^2} [1 - \Phi_1(R)] \hat{\mathbf{R}}.
\end{equation}

\noindent Obviously this is far from the force given in (\ref{force}).\\
\indent If, however, we define an ``area'' we will call the \textit{Schwarzschild area}, $A_S$, as $4 \pi(2GM/c^2)^2$, which has no relation to the area of the black hole, we find, using $M = c^2 \sqrt{A_S}/4\sqrt{\pi} G$,
\begin{equation}
\frac{\dif \mathcal{S}}{\dif A_S} = \frac{1}{8 \pi l_P^2} \left[1 + \Phi_1(\sqrt{A_S}/2\sqrt{\pi}) + \sqrt{A_S} \frac{\dif \Phi_1 (\sqrt{A_S}/2 \sqrt{\pi})}{\dif \sqrt{A_S}} \right].
\end{equation}

\noindent Since we have used the formula from the pure Newtonian calculation which has an extra factor, we expect the force to be (as in our companion article),
\begin{equation}
\mathbf{F} = - \frac{GMm}{R^2} 8 \pi l_P^2 \frac{\dif \mathcal{S}}{\dif A_S} \hat{\mathbf{R}},
\end{equation}

\noindent and
\begin{equation}
\mathbf{F} = -\frac{GMm}{R^2} \left[1 + \Phi_1(R) + R \frac{\dif \Phi_1(R)}{\dif R} \right] \hat{\mathbf{R}}.
\end{equation}

\noindent While this result is almost of the form of $\mathbf{F}$ of Eq. (\ref{force}), the sign of the last term in brackets is wrong, so in neither of these cases does the circle close. While the circle closes in ordinary Newtonian theory, it does not in our cases.\\
\indent This conundrum could be due to several factors, for example, a straightforward application of the formula of Verlinde to generate the force $\mathbf{F}$ or maybe the use of the simple relation between $T_{MLH}$ and $\mathcal{S}$. It is difficult to see how to modify Verlinde's formula for a more complex Newtonian theory. In our companion article a similar mismatch occurs when we begin with a black hole solution to a specific modified Newtonian theory. There the force is already known, and the force derived from applying the Verlinde formula also does not match.

\section{Conclusion}

\noindent If gravity is not a fundamental force, as previously suggested \cite{Jacobson1, Jacobson2, Verlinde}, then avenue opens which can lead to new concepts and paradigms. Searching for new effects through terms that correct some properties might give us an opportunity to find such effects. In the present work we use simple arguments in order to show how generalized entropies may give the corrections we are looking for, corrections to Newtonian forces,  potentials, and to the Hawking temperature.\\
\indent As we discussed in the main text, in the case of the quantum entropy we are able to interpret such correction terms as coming from the weak field approximation to a (still unknown) generalized gravity with further terms due to the quantum corrections to the entropy. If this entropic concept happens to be true, the ``weak field'' approximation for such an effective quantum gravity has to have as classical limit the Newtonian law we have just found. Although we can interpret these terms as effective matter densities, modifications to the Laplace operator seems perhaps to be a better idea. While the ad hoc equations (\ref{laplacian1}) and (\ref{laplacian2}) are not at all promising, we may be able to use modifications of the ideas of Jacobson et al. \cite{Jacobson3, Jacobson4}. We are pursuing the matter at present.\\
\indent As a concluding note, modifications to the one-quarter-of-area formula for the entropy were considered previously in \cite{Visser, JKM}, for example, by means of the Noether charge approach \cite{Wald} in modified theories of gravity. These authors consider Newton's constant explicitly in the Bekenstein-Hawking $\mathcal{S}_{BH}$, writing\footnote{In this formula $\hbar = c = 1$.} $\mathcal{S}_{BH} = A/4G$.  For modified theories they write $\mathcal{S}_{BH} = A/4G_N$, $G_N$ a modified (or effective) Newton's constant. Of course, the expressions for our entropies can be written in this form with $G_N$ a function of $A$.  For the same modified theories the author in \cite{Vollik} found corrections to $G_N$, using the Wald method. The terms appearing in such modifications are evaluated at the (Killing) horizon of the black hole. Remember that in Verlinde's derivation of Newton's law, this horizon is the area of the holographic sphere, which in this article is the sphere with the Michell-Laplace radius. The Newtonian force, as well as the potential, are local quantities evaluated on this sphere, that is, at the horizon in Verlinde's approach.

{\bf Acknowledgments} We'd like to thank M. Sabido for fruitful discussions, and C.A. L\'opez Castro for his enthusiasm in learning these topics. The work of A. Mart\'inez-Merino is supported by a CONACyT post-doctoral grant. O. Obreg\'on was supported by CONACyT Project numbers 257919 and 258982, PROMEP and UG Projects.

\bibliographystyle{apsrev4-1}

\end{document}